\documentclass{emulateapj}

\shorttitle{Evolution of IMXBs driven by MB of Ap/Bp stars: I. UCXBs}
\shortauthors{Chen \& Podsiadlowski}

\begin{document}


\title{Evolution of intermediate-mass X-ray binaries driven by magnetic braking of Ap/Bp stars: I. ultracompact X-ray binaries }


\author{Wen-Cong Chen$^{1,2}$,  and Philipp Podsiadlowski$^{2}$}
\affil{$^1$ School of Physics and Electrical Information, Shangqiu Normal University, Shangqiu 476000, China;\\
$^2$ Department of Physics, University of Oxford, Oxford OX1 3RH, UK;
chenwc@pku.edu.cn}



\begin{abstract}
It is generally believed that Ultracompact X-ray binaries (UCXBs)
evolved from binaries consisting of a neutron star accreting from a
low-mass white dwarf or helium star where mass transfer is driven by
gravitational radiation. However, the standard white-dwarf
evolutionary channel cannot produce the relatively long-period ($40 -
60$\,min) UCXBs with high time-averaged mass-transfer rate. In this
work, we explore an alternative evolutionary route toward UCXBs where
the companions evolve from intermediate-mass Ap/Bp stars with an
anomalously strong magnetic field ($100 - 10000$\,G).  Including the
magnetic braking caused by the coupling between the magnetic field and
an irradiation-driven wind induced by the X-ray flux from the
accreting component, we show that intermediate-mass X-ray binaries
(IMXBs) can evolve into UCXBs. Using the \emph{MESA} code, we have
calculated evolutionary sequences for a large number of IMXBs.  The
simulated results indicate that, for a small wind-driving efficiency
$f=10^{-5}$, the anomalous magnetic braking can drive IMXBs to an
ultra-short period of 11 min. Comparing our simulated results
with the observed parameters of fifteen identified UCXBs, the
anomalous magnetic braking evolutionary channel can account for the
formation of seven and eight sources with $f=10^{-3}$, and $10^{-5}$,
respectively. In particular, a relatively large value of $f$ can fit
three of the long-period, persistent sources with high mass-transfer
rate. Though the proportion of Ap/Bp stars in intermediate-mass stars
is only 5\%, the lifetime of the UCXB phase is $\ga$ 2 Gyr, producing
a relatively high number of observable systems, making this an
alternative evolutionary channel for the formation of UCXBs.
\end{abstract}


\section{Introduction}
Ultracompact X-ray binaries (UCXBs) are a sub-population of low-mass
X-ray binaries (LMXBs) with ultra-short orbital periods (usually
less than 1 hour). They are thought to be accretion-powered X-ray
sources, in which a neutron star (NS) or black hole (BH, though BH
accretors have not yet been detected in UCXBs) accretes matter from a
donor star by Roche-lobe overflow \citep{savo86}. From the orbits of
the UCXBs, their donor stars can be constrained to be partially or fully
degenerate stars such as white dwarfs (WDs) or helium (He) stars
\citep{rapp82,nels86,pods02,delo03}. In some UCXBs or UCXB candidates,
the spectra imply that the transferred material is either mainly
composed of He or carbon and oxygen, which provides support for the
donor identifications \citep{nele04,nele06}. So far, there are fifteen
identified UCXBs, ten persistent sources and five transient sources.

It is worth pointing out the broad astrophysical significance of
UCXBs.  They offer important information on stellar and binary
evolution, constraining the accretion process, angular-momentum loss
mechanisms and the common-envelope phase.  In addition, UCXBs are
thought to be possible gravitational-wave sources, which can be
detected by LISA \citep{nele09}.

In globular clusters, it is generally accepted that UCXBs originated
from dynamic processes such as direct collisions
\citep{verb87,rasi91,davi92,ivan05,lomb06,ivan10}, tidal captures
\citep{bail87,pods02}, and exchange interactions
\citep{davi98,rasi00,ivan10}. In the Galactic field, these dynamical
process can safely be ignored.  In the field, the donor stars in the
progenitor systems of UCXBs include two cases: main-sequence stars,
and compact WDs or He stars. In standard low/intermediate-mass X-ray
binaries (L/IMXBs) with a main-sequence donor star, magnetic braking
can drive the binaries to a relatively short period of $2 - 3$ hours.
When the donor star becomes fully convective, magnetic braking ceases
and gravitational radiation becomes the dominant angular-momentum loss
mechanism, possibly leading to the formation of a UCXB. For a binary
system consisting of a NS and a (sub)giant, mass transfer may be
dynamically unstable. This leads to a common-envelope (CE) phase and
ultimately the formation of NS + WD or NS + He star binaries with
compact orbits.  Gravitational radiation will make the orbits shrink
till the WDs or He stars fill their Roche lobes and start to
transfer material to the NSs. With the further decay of the orbits,
the binaries will now appear to be UCXBs
\citep{tutu93,iben95,yung02,belc04,haaf12a}.

At present, there exist some intriguing problems in the formation of
UCXBs. First, a very narrow range in initial parameters cannot explain
the relatively large number of detected UCXBs \citep{sluy05a}. Second,
the evolution towards UCXBs strongly depends on the efficiency of
magnetic braking. It is possible that the standard magnetic-braking
model overestimates the angular-momentum loss rate (also see section
3.2). \cite{sluy05b} found that, it is difficult to form UCXBs within a
Hubble time under the modified magnetic-braking law given by
\cite{quel98} and \cite{sill00}. Third, \cite{haaf12b} found that the
time-averaged luminosity for UCXBs with periods longer than 50 min are
significantly higher than the predictions from the theoretical mass-transfer
rate. Recently, \cite{hein13} argued that the higher average
mass-transfer rates in three persistent UCXBs with relatively long
orbital periods ($40 - 60$ min) cannot be produced by the standard WD
evolutionary scenario.

NS + He stars binary can produce the group of UCXBs with long orbital
periods ($40 - 60$ min) and high mass-transfer rates
\citep{hein13}. If one includes the tidal torque between the binary
and a hypothetical circumbinary disk, this population of UCXBs can
evolve from LMXBs if a relatively high fraction ($\delta \ga 0.008$)
of the mass transferred is fed to the circumbinary disk
\citep{ma09b}. However, it is difficult to reproduce the relatively high
birth rates for the above two models because of the short lifetime
($\sim 10^{7}$ yr) in the UCXB stage.

To solve some of these problems, a new evolutionary channel to form
UCXBs may be required or a more efficient angular-momentum loss
mechanism. In this work, we present an alternative evolutionary
channel for UCXBs.  In this channel we assume that some
intermediate-mass donor stars have an anomalously strong magnetic
field ($10-10000$ G) and that the magnetic braking is produced by the coupling
between the magnetic field and the irradiation-driven wind and show
that this can convert IMXBs into UCXBs.

\section{Description of binary evolution}

\subsection{The binary evolution code}
In this work, we calculate the evolution of IMXBs using version
7624 of the Modules for Experiments in Stellar Astrophysics code
\citep[\texttt{MESA; }][]{paxt11,paxt13,paxt15}. In particular,
\texttt{MESAbinary} is a \texttt{MESA} module that can evolve a binary
system including a normal star and a point-mass companion or another
star. We start with binary systems consisting of an
intermediate-mass donor star (with a mass of $M_{\rm d} = 1.6 - 5.0
~\rm M_{\odot}$) and a NS (with a mass of $M_{\rm NS} = 1.4 ~\rm
M_{\odot}$) in a circular orbit. For the donor star, solar
composition ($X = 0.70, Y = 0.28$, and $Z = 0.02$) is adopted. The
effective Roche-lobe radius of the donor star is given by \cite{egg83}
\begin{equation}
\frac{R_{\rm L}}{a}=\frac{0.49q^{2/3}}{0.6q^{2/3}+ {\rm
ln}(1+q^{1/3})},
\end{equation}
where $a$ is the orbital separation, and $q = M_{\rm d}/M_{\rm NS}$ is the
mass ratio of the binary.

\subsection{Mass transfer}
When the donor star evolves to fill its Roche lobe, the material is
transferred to the NS at a rate of $\dot{M}_{\rm tr}$ \textbf{($<0$)} via Roche-lobe
overflow. \texttt{MESAbinary} offers two choices for the mass-transfer
schemes \citep{paxt15}.  Here we use the default binary control
parameters, which utilize the Ritter scheme \citep{ritt88}. During the
mass exchange, the accretion process onto the NS is limited to the
Eddington accretion rate
\begin{eqnarray}
\dot{M}_{\rm Edd}=1.8\times10^{-8}\left(\frac{M_{\rm NS}}{1.4\rm M_{\odot}}\right)\left(\frac{0.2}{\eta}\right) \nonumber \\
\left(\frac{1.7}{1+X}\right)\,\rm M_{\odot}\,yr^{-1},
\end{eqnarray}
where $\eta=GM_{\rm NS}/(R_{\rm NS}c^{2})$ is the energy conversion
efficiency ($G$ is the gravitational constant, $R_{\rm NS}$ the NS
radius, and $c$ the speed of light in vacuo). In this work, a constant
NS radius of $R_{\rm NS}=10^{6}~\rm cm$ is adopted, and $X$ is the
hydrogen abundance in the accreting material. Similar to
\cite{pods02}, the accretion efficiency of the NS is assumed to be
0.5; thus, the fraction of transferred mass that is lost from
  the vicinity of the NS is
\begin{equation}
\beta= \left\{\begin{array}{l@{\quad}l} 0.5,
& -\dot{M}_{\rm tr}\leq 2\dot{M}_{\rm Edd} \strut\\
\frac{\dot{M}_{\rm tr}+\dot{M}_{\rm Edd}}{\dot{M}_{\rm tr}}
, &  -\dot{M}_{\rm tr}\geq 2\dot{M}_{\rm Edd}. \strut\\\end{array}\right.
\end{equation}
The accretion rate of the NS can then be written as
\begin{equation}
\dot{M}_{\rm NS}=  -(1-\beta)\dot{M}_{\rm tr}.
\end{equation}

In addition, we also consider irradiation-driven winds. In a compact
binary, the stellar wind from the donor star induced by the X-ray
radiation cannot be ignored during the accretion of the NS
\citep{rude89,tava93}. \cite{haaf13} found that irradiation of the
donor star plays a vital role in forming UCXBs with an orbital period
longer than 40 min. A fraction of the X-ray luminosity ($L_{\rm
  X}=\eta\dot{M}_{\rm NS}c^{2}$) that the donor star receives is
assumed to drive a wind from the surface of the donor star and converted
into the kinetic energy of the wind (with a velocity given by the escape
speed from the donor's surface). Hence, the stellar wind-loss rate
satisfies the relation
\begin{equation}
L_{\rm X}\frac{\pi R_{\rm d}^{2}}{4\pi a^{2}}f=-\frac{GM_{\rm d}\dot{M}_{\rm wind}}{R_{\rm d}},
\end{equation}
where $R_{\rm d}$ is the radius of the donor star, $f$ is the
wind-driving efficiency. The mass-loss rate of the donor star
$\dot{M}_{\rm d}=\dot{M}_{\rm tr}+\dot{M}_{\rm wind}.$

\subsection{Orbital angular-momentum loss mechanisms}
The loss of orbital angular momentum plays a vital role during the
evolution of binary systems.  In this work, the rate of orbital-angular
momentum loss of IMXBs consists of three terms
\begin{equation}
\dot{J}=\dot{J}_{\rm GR} + \dot{J}_{\rm ML} +\dot{J}_{\rm MB}.
\end{equation}
In equation (6), the first term $\dot{J}_{\rm GR}$ is
produced by gravitational-wave radiation, given by \citep{land71}
\begin{equation}
\dot{J}_{\rm GR}=-\frac{32G^{7/2}}{5c^{5}}\frac{M_{\rm NS}^{2}M_{\rm d}^{2}(M_{\rm NS}+M_{\rm d})^{1/2}}{a^{7/2}}.
\end{equation}

The second term in equation (6) takes into account the orbital
angular-momentum loss due to the systemic mass loss. It includes
the mass loss from the irradiation-driven wind of the
donor star and the mass outflow from the accreting NS. The
former is thought to carry away the specific orbital angular momentum
of the donor star. The latter should form an isotropic wind in the
vicinity of the NS and is then ejected with the specific orbital
angular momentum of the NS. Thus the angular-momentum loss rate becomes
\begin{equation}
\dot{J}_{\rm ML}=\frac{2\pi a^{2}}{(M_{\rm NS}+M_{\rm d})^{2}P_{\rm orb}}(M_{\rm NS}^{2}\dot{M}_{\rm wind}+\beta M_{\rm d}^{2}\dot{M}_{\rm tr}),
\end{equation}
where $P_{\rm orb}$ is the orbital period of the binary.

The third term in equation (6) specifies the angular-momentum loss
rate caused by magnetic braking, which plays a key role leading to the
formation of UCXBs in our model. We have already accounted for the
direct influence of the irradiation-driven winds on the orbital
evolution in $\dot{J}_{\rm ML}$. However, its indirect effect is much
more important if the magnetic field is strong.  It is generally
thought that the stellar wind always couples to the stellar magnetic
field up to the magnetospheric radius ($r_{\rm m}$) beyond which the
magnetic field can no longer force the stellar wind to
co-rotate. Because the wind material is tied to the magnetic field
lines up to $r_{\rm m}$, its specific angular momentum
is considerable larger than the specific angular momentum of the binary
system \citep{verb81}. Assuming that the irradiation-driven winds
are expelled at the magnetospheric radius, the loss rate of angular-momentum
loss produced by
magnetic braking can be written as \citep{just06}
\begin{eqnarray}
\dot{J}_{\rm MB}&=&\dot{M}_{\rm wind}r_{\rm m}^{2}\frac{2\pi}{P_{\rm orb}} \nonumber\\
&=&-B_{\rm s}R_{\rm d}^{13/4}\sqrt{-\dot{M}_{\rm wind}}(GM_{\rm d})^{-1/4}\frac{2\pi}{P_{\rm orb}},
\end{eqnarray}
where $B_{\rm s}$ is the surface magnetic field of the donor
star \footnote{Equation (9) is based on a simple analytical model for the
physical process of magnetic braking. We can not use the model given by
\cite{verb81} which is an empirical model to reproduce
observed spin rates of low-mass stars, which is not directly applicative to
Ap/Bp stars with their relatively strong magnetic fields.}. Meanwhile, the donor star in a close binary should be tidally
locked, which would continuously spin up (or spin down) the donor star
till its spin is synchronized with the orbital motion
\citep{patt84}. This implies that the angular momentum carried away in
the magnetic wind is ultimately drawn from the orbital angular momentum
of the binary, making magnetic braking an important
orbital angular-momentum loss mechanism. From equations
(5) and (9), and combining $L_{\rm
  X}=\eta\dot{M}_{\rm NS}c^{2}$ and Kepler's third law, the orbital angular-momentum loss rate due to magnetic braking
can be found to be
\begin{equation}
\dot{J}_{\rm MB}=-\frac{B_{\rm s}c}{2}\sqrt{\frac{f\eta\dot{M}_{\rm NS}(M_{\rm NS}+M_{\rm d})}{a^{5}}}\left(\frac{R_{\rm d}^{19}}{GM_{\rm d}^{3}}\right)^{1/4}.
\end{equation}
If the donor stars lose their radiative core, magnetic braking is assumed to cease \citep{rapp83,spru83}.

\begin{figure*}
\centering
\begin{tabular}{ccc}
\includegraphics[width=0.33\textwidth,trim={10 0 0 30},clip]{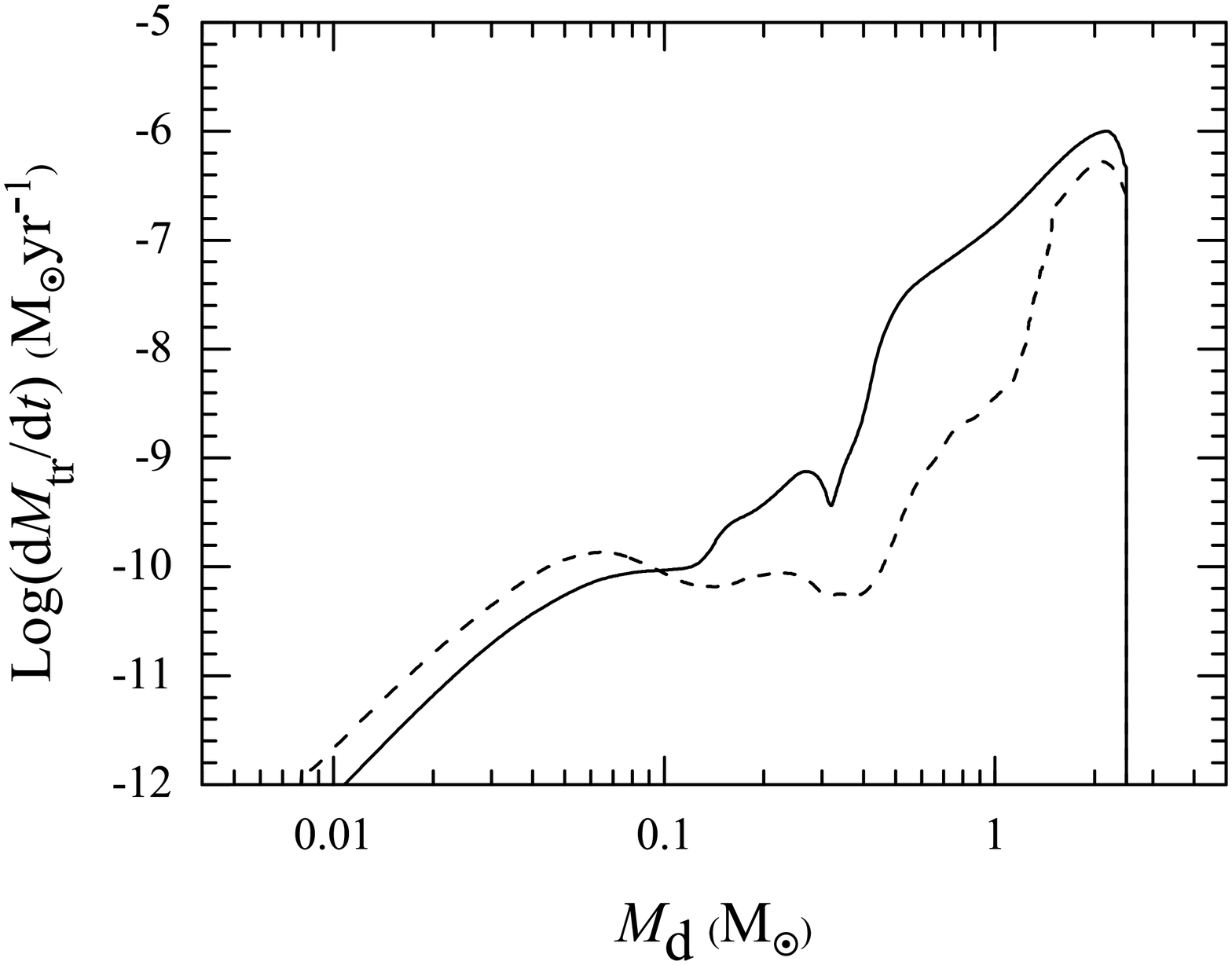} &    \includegraphics[width=0.33\textwidth,trim={0 0 0 30},clip]{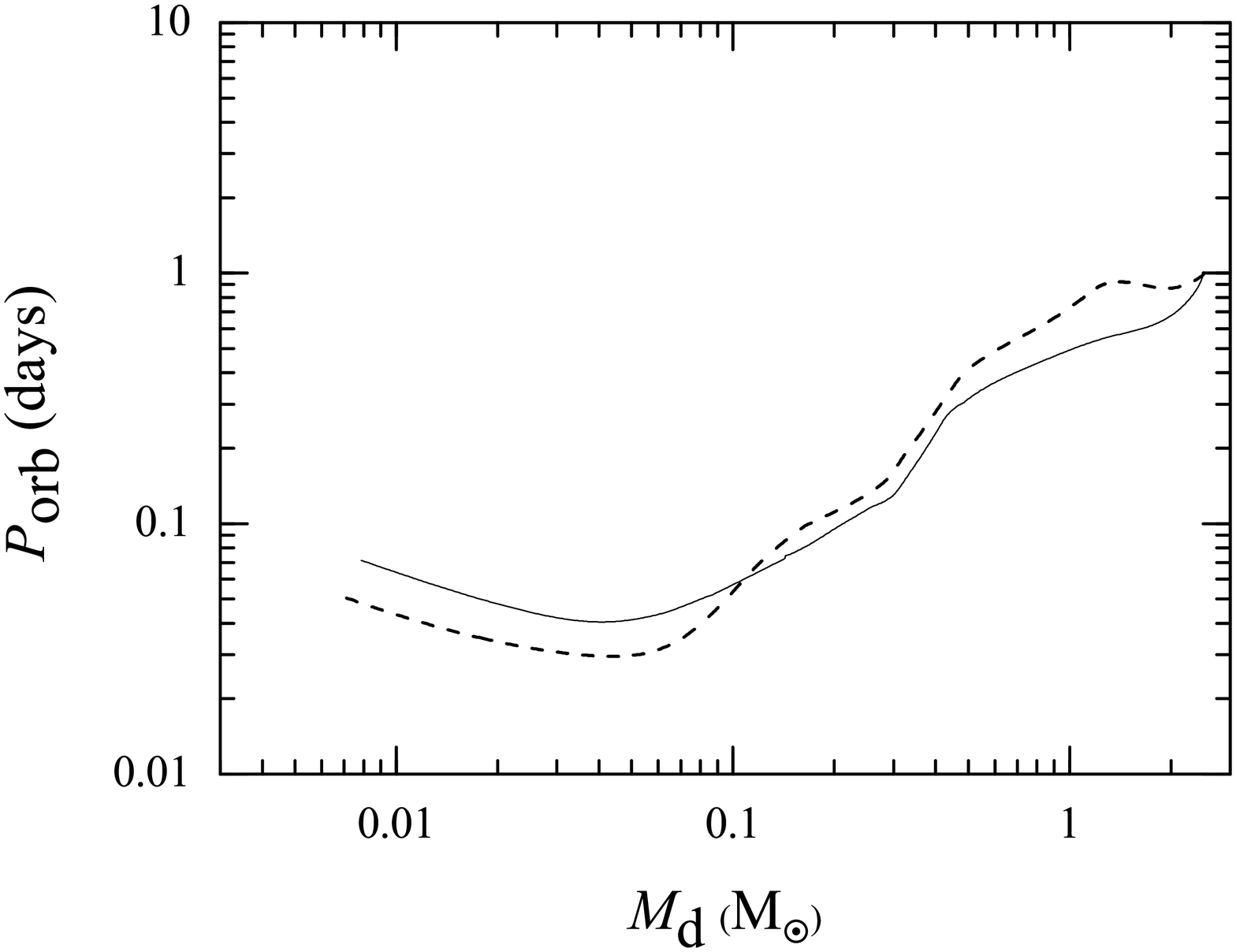} &\includegraphics[width=0.33\textwidth,trim={0 0 0 30},clip]{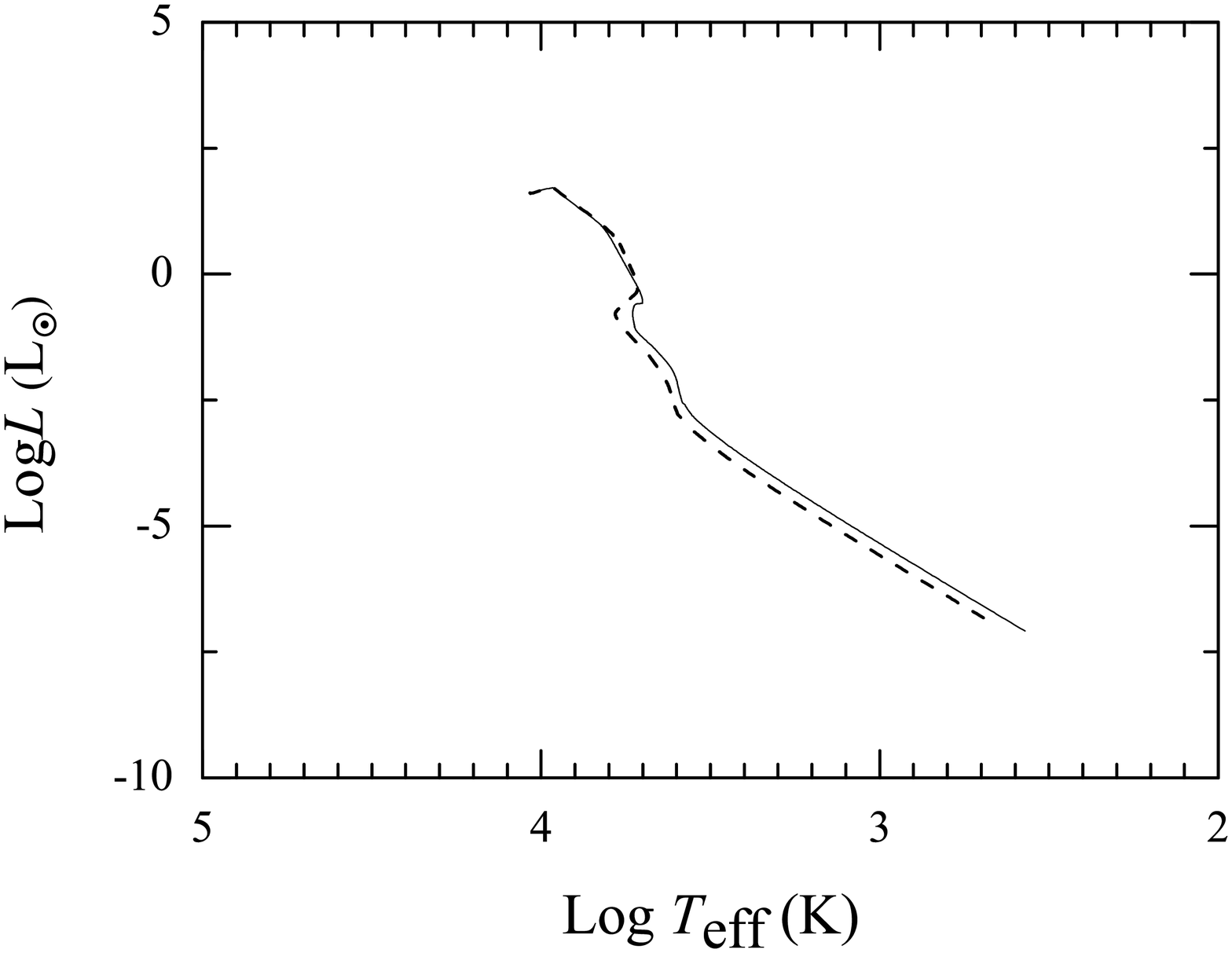} \\
\end{tabular}
\caption{\label{fig:pulses}Mass-transfer rate (left panel) and orbital
  period (middle panel) as a function of the donor-star mass, HR
  diagram (right panel) of the donor star when the initial donor-star
  mass is $2.5~\rm M_{\odot}$ and the initial orbital period $P_{\rm
    orb,i} = 1.0$ d. The wind-driving efficiency
   $f = 10^{-3}$. The solid, and dashed curves represent the
  anomalous magnetic braking model with $B_{\rm s}$ = 1000 G, and 100
  G, respectively.}
\end{figure*}

\section{Results}
\subsection{Wind-driving efficiency $f = 10^{-3}$}
Because $\dot{J}_{\rm mb}\propto B_{\rm s}f^{1/2}$, there is
    a degeneracy between $B_{\rm s}$ and $f$ in equation (10), and we
    can keep one fixed while we vary the other.
  Similar to \cite{just06}, in
  this subsection, we take a relatively large wind-driving efficiency
  $f = 10^{-3}$. To investigate the influence of the surface magnetic
  field on the evolution, we first compare in Figure~1 the
  evolutionary results for magnetic fields of 100 G and of 1000 G,
  respectively, for an IMXB with a donor star of 2.5 $\rm M_{\odot}$
  and an initial period of 1 d. As shown in this figure, a strong
  magnetic field can result in a very efficient angular-momentum loss
  rate and rapid mass transfer in most evolutionary stages, as
  expected from equation (10). Interestingly, the system with the
  weaker magnetic field has a lower minimum period (see also the
  discussion in Section 4).  For simplicity, we always take a 1000 G
  surface magnetic field in this work, and examine the influence of
  the wind-driving efficiency $f$ on the formation of UCXBs.

\begin{figure*}
\centering
\begin{tabular}{cc}
\includegraphics[width=0.4\textwidth,trim={30 10 30 30},clip]{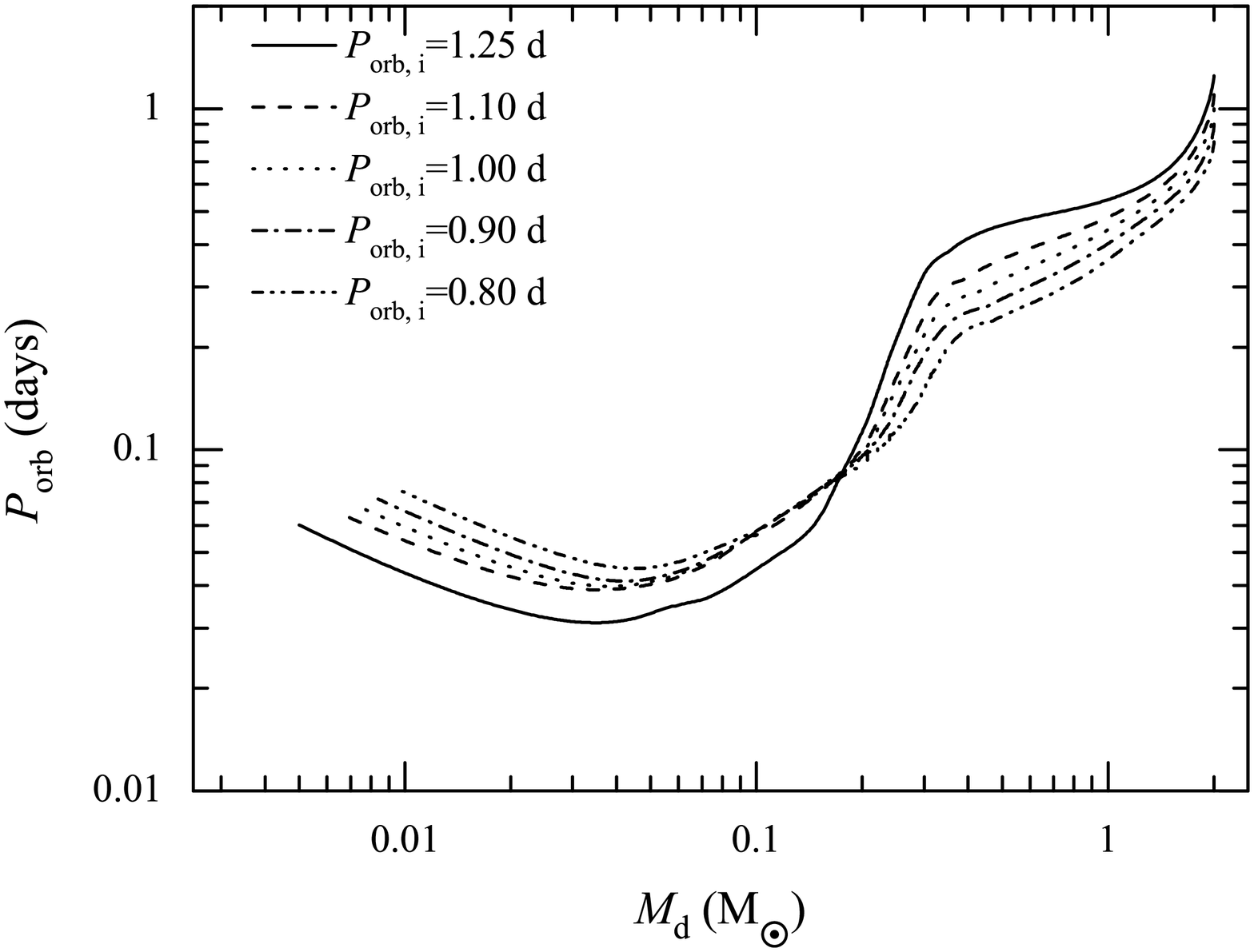} &    \includegraphics[width=0.4\textwidth,trim={30 10 30 30},clip]{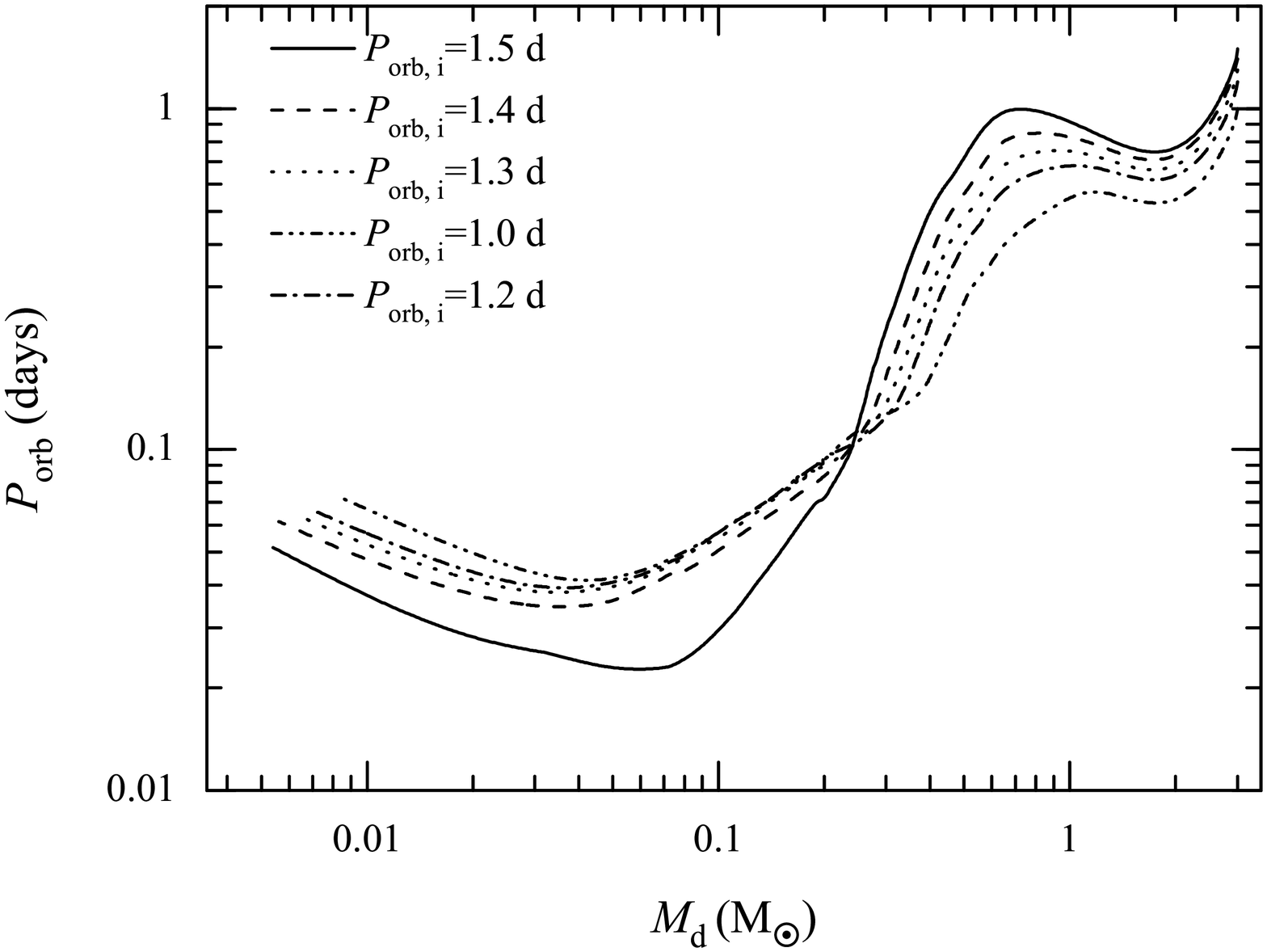} \\
\end{tabular}
\caption{\label{fig:pulses}Evolutionary tracks of binary systems with
  different initial donor-star masses and initial orbital periods in
  the $P_{\rm orb} - M_{ \rm d}$ diagram. The wind-driving efficiency
   $f = 10^{-3}$. The left and right panels
  denote the evolution of IMXBs with donor stars of 2.0, and $3.0~\rm
  M_{\odot}$, respectively.}
\end{figure*}

To understand the evolutionary history of UCXBs originating from this
anomalous magnetic-braking model with irradiation, we plot the
evolution of IMXBs with two initial donor-star masses and different
initial orbital periods in the $P_{\rm orb} - M_{\rm d}$ plane (see
also Figure 2). In order to evolve towards ultra-short orbital
periods, the initial orbital period of the IMXBs should be near the
so-called bifurcation period of 1.25 d and 1.5 d, depending on
the initial donor-star masses. The bifurcation period is defined as
the longest initial orbital period that forms UCXBs within a Hubble
time \citep{sluy05a,sluy05b}; it strongly depends on the
magnetic-braking efficiency and mass loss from the binary systems
\citep{pyly88,pyly89,ergm98,pods02,ma09a}. The minimum periods are
very sensitive to the initial periods: as shown in Figure 2, higher
initial periods tend to lead to lower minimum periods. In Table 1, we
present some of the relevant evolutionary quantities of IMXBs with
their respective bifurcation periods for different initial donor-star
masses. Table 1 indicates that these donor stars spend a long
time of their evolution where H in the center is almost exhausted, in
agreement with the results given by \cite{nels86} and \cite{fedo89}. High He
abundances in the core result in a more compact donor star and
correspondingly shorter orbital periods \citep{tutu87,lin11}. Compared
to the previous models, the anomalous magnetic braking increases the
bifurcation period because of the more efficient angular momentum loss.

\begin{figure*}
\centering
\begin{tabular}{cc}
\includegraphics[width=0.4\textwidth,trim={30 50 80 10},clip]{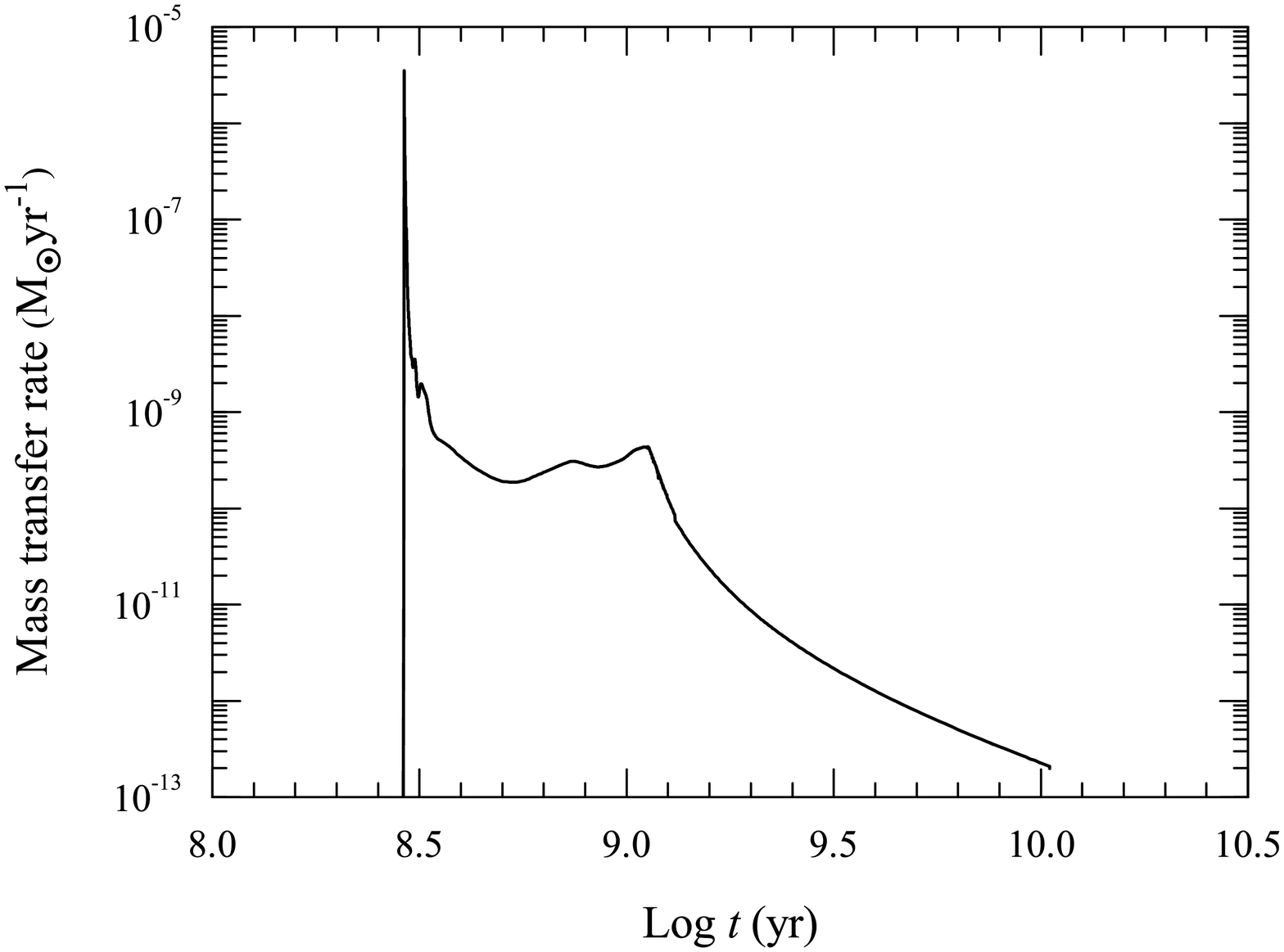} &    \includegraphics[width=0.4\textwidth,trim={30 50 80 10},clip]{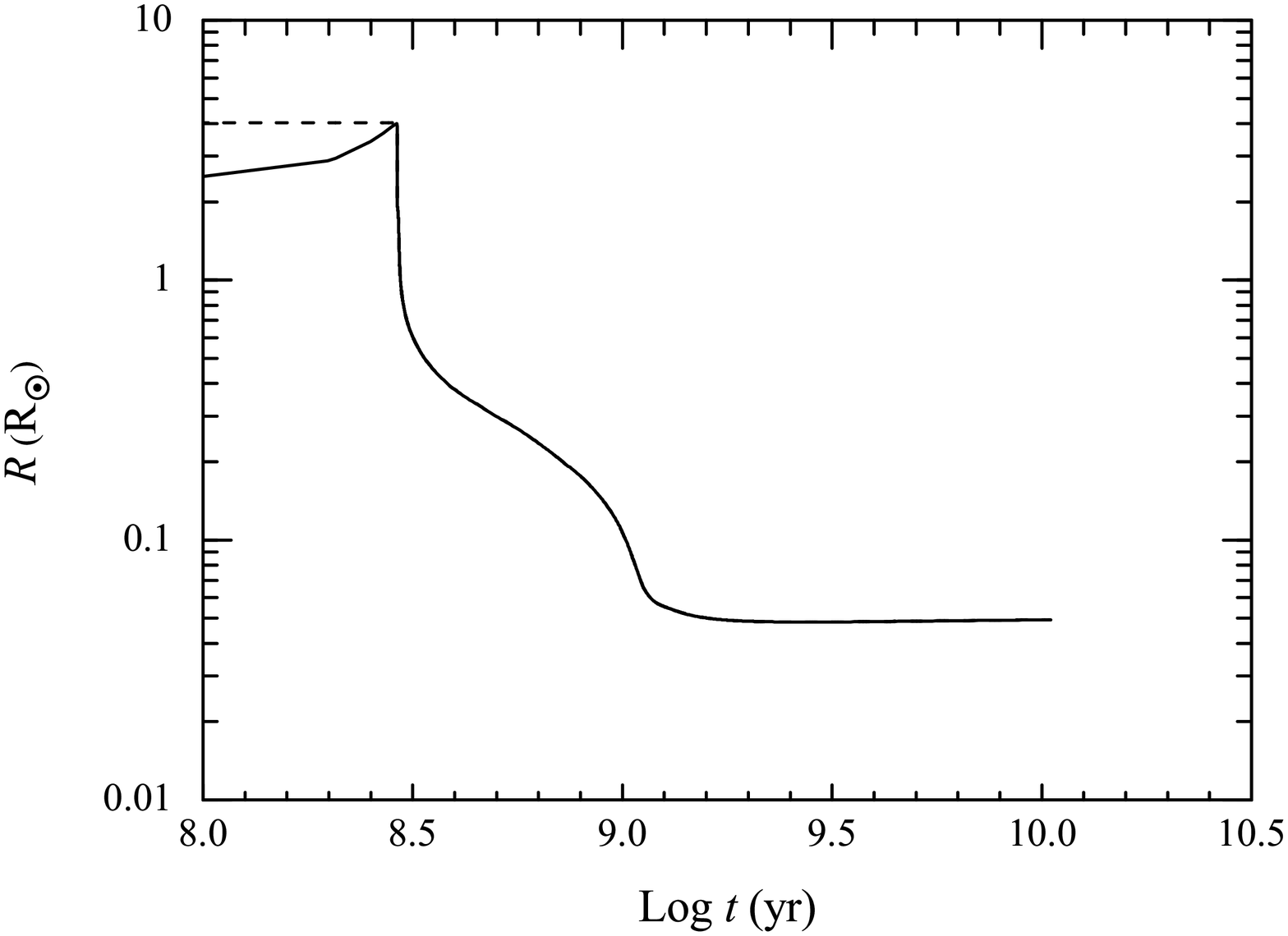} \\
\includegraphics[width=0.4\textwidth,trim={30 10 80 10},clip]{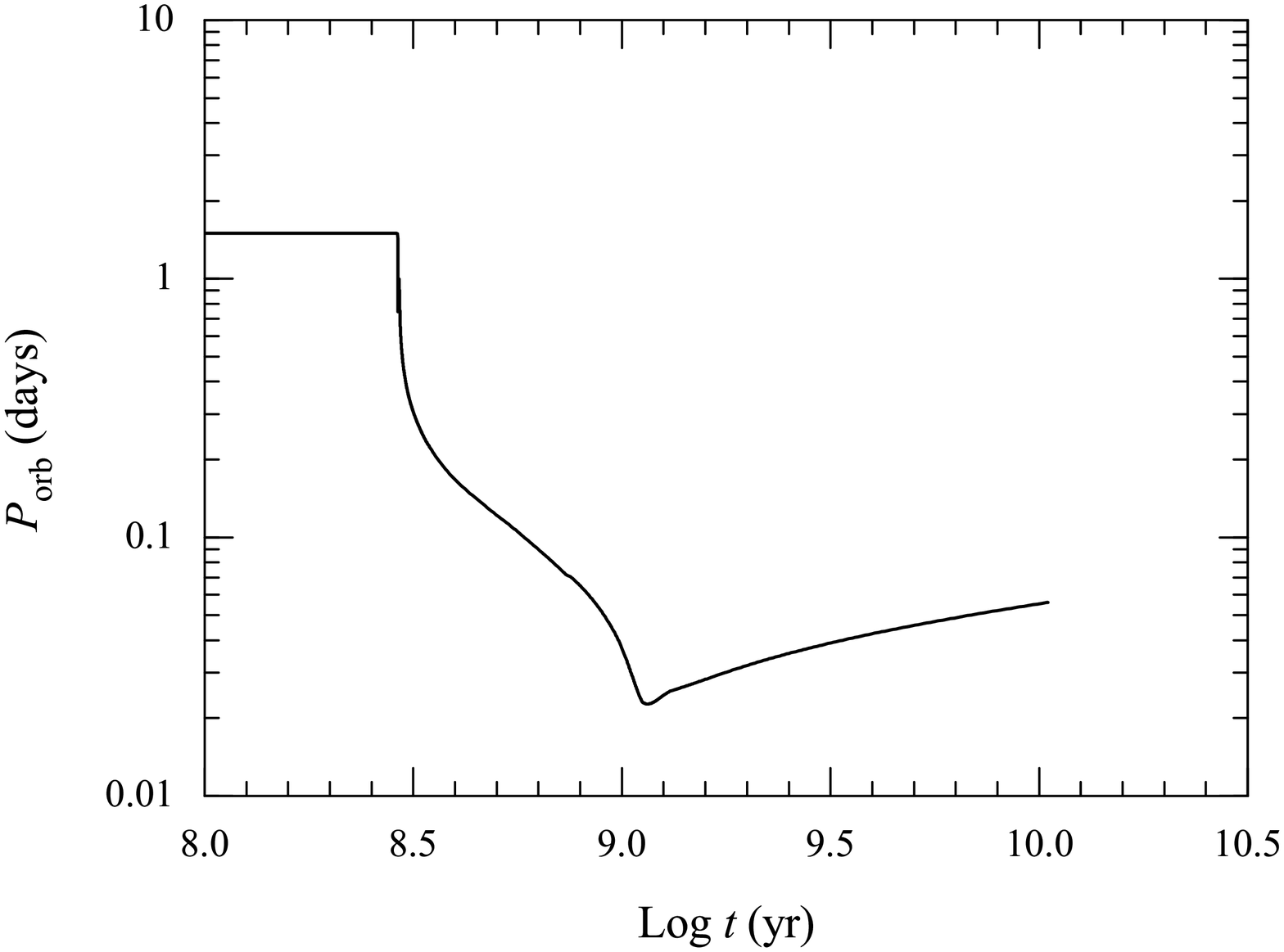} &   \includegraphics[width=0.4\textwidth,trim={30 10 80 10},clip]{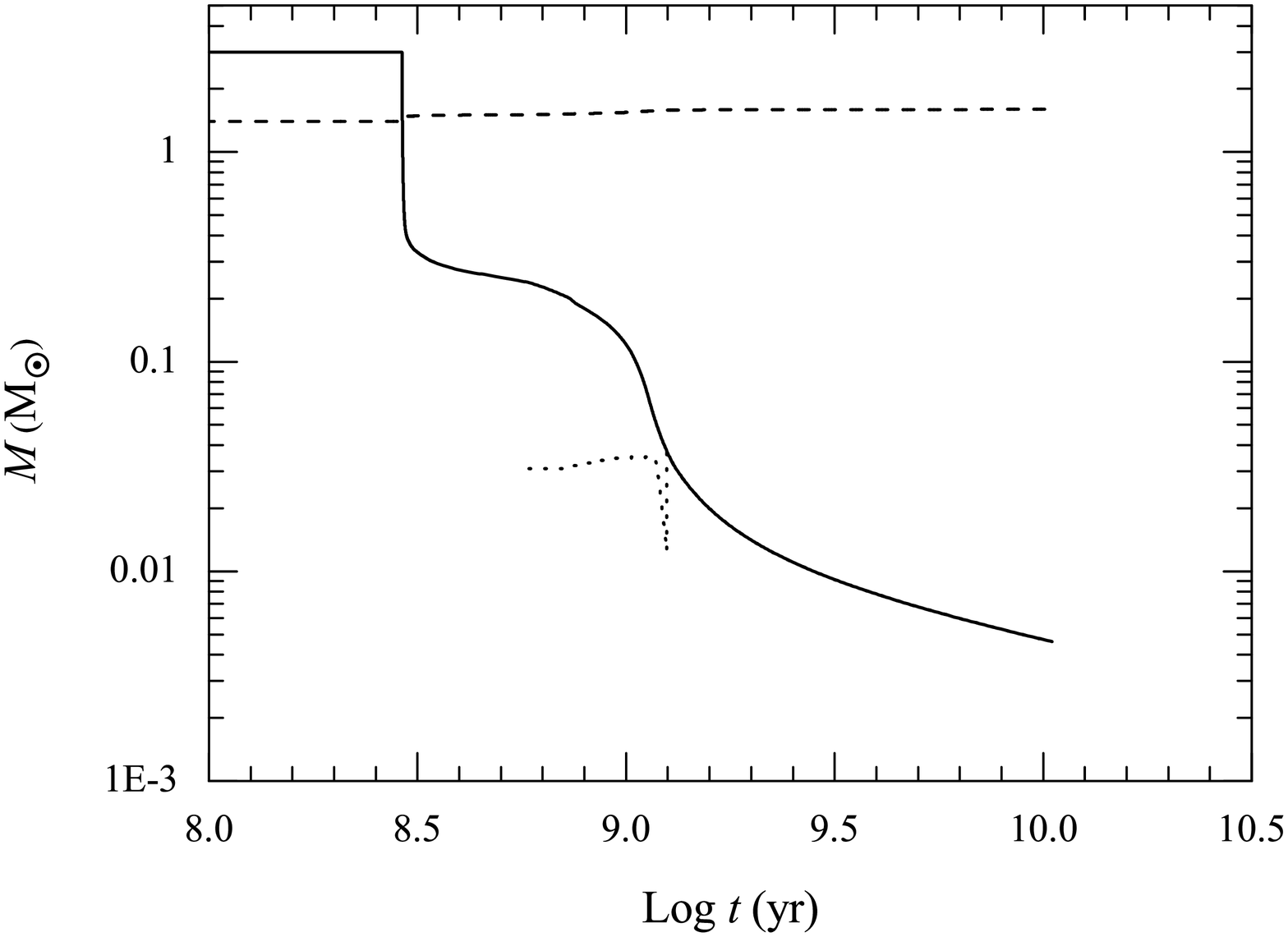} \\
\end{tabular}
\caption{\label{fig:pulses}Evolution of the main binary parameters as a function of the donor star age from the zero-age main sequence for an IMXB with a donor-star mass of $3~\rm M_{\odot}$ and an initial
  period of 1.5 days. The wind-driving efficiency
   $f = 10^{-3}$. Top left panel: mass-transfer rate. Bottom left
  panel: orbital period. Top right panel: radius (solid curve) and
  Roche-lobe radius (dashed curve) of the donor star. Bottom right
  panel: mass of the donor star (solid curve), NS (dashed curve), and
  the core (dotted curve).}
\end{figure*}

Figure 3 illustrates the evolutionary sequence for an IMXB with a 3
$\rm M_{\odot}$ donor star and an initial orbital period of 1.5
d. After 0.29 Gyr of nuclear evolution, the donor star starts to fill its
Roche lobe when hydrogen in the center is almost exhausted (at this
point $X_{\rm c}=0.0436$, and $Y_{\rm c}=0.937$, i.e.\ the donor star
is near the end of the main sequence). Because the material is
transferred from the more massive donor star to the less massive NS,
mass transfer first occurs on the thermal timescale of the donor at a
rate $10^{-7} - 10^{-6}\, \rm M_{\odot}yr^{-1}$. Once the mass ratio
is less than 1, the mass-transfer rate changes to a sub-Eddington
rates. From $t \sim$ 0.3 to 1 Gyr, the mass-transfer rate is in the
range of $10^{-10} - 10^{-9}\, \rm M_{\odot}yr^{-1}$, and then sharply
decreases to a very low value. At $P_{\rm orb}= 1.7 $ hour, the
angular-momentum loss rate due to gravitational radiation begins to
exceed that due to magnetic braking. The orbital period of the binary continues to
decrease and reaches a period minimum of 32 min when $t =$ 1.15
Gyr. At some point, magnetic braking turns off because the radiative
core vanishes (at this point the remaining donor star has a low-mass
He core). At $t =$ 0.97 Gyr, the IMXB enters a phase where the period
enters the UCXB regime (with decreasing period evolution) and stays in
this regime until $t =$ 3.8 Gyr, when its orbital period exceeds 1
hour again. Therefore, our simulation show that the active UCXB
lifetime in the period-decreasing and period-increasing phases are
$\sim0.2$ Gyr, and $\sim2.6$ Gyr, respectively.

\begin{table}
\begin{center}
\caption{Selected evolutionary properties for IMXBs with different
  initial donor-star masses. We list the initial donor-star mass,
  bifurcation period, the orbital period at the beginning of Roche
  lobe overflow, the central H and He abundance at the beginning of
  Roche-lobe overflow, the minimum period, the donor star mass and the
  mass-transfer rate at the minimum period. The wind-driving efficiency
   $f = 10^{-3}$.\label{tbl-2}}
\begin{tabular}{@{}llllllll@{}}
\hline\hline\noalign{\smallskip}
$M_{\rm d,i}$ & $P_{\rm bif}$ &  $P_{\rm rlov}$ &$X_{\rm c,rlov}$ &$Y_{\rm c,rlov}$ &$P_{\rm min}$ & $M_{\rm d, min}$ & ${\rm log}\dot{M}_{\rm d, min}$ \\
 ($\rm M_{\odot}$)     &  (d)  & (d)   &   &           &    (min) & ($\rm M_{\odot}$ ) & ($\rm M_{\odot}yr^{-1}$)      \\
\hline\noalign{\smallskip}
1.6 & 1.03&1.01& 0.05&0.93& 43& 0.035& $-10.2$\\
1.8 & 1.15  &1.09 & 0.08& 0.90&49&0.036& $-10.3$ \\
2.0 & 1.27 &1.21&0.06&0.92& 44 &0.036  & $-10.2$ \\
2.2 & 1.33 &1.31&0.04&0.94&40&0.034   & $-10.2 $ \\
2.4 & 1.36 & 1.31  &0.06&0.92&44&0.036& $-10.2 $  \\
2.6 & 1.42 & 1.41 & 0.05 & 0.93 & 40 &0.034& $-10.2$ \\
2.8 & 1.44  &1.43 &0.06&0.92& 38&0.035 & $-10.2$ \\
3.0 & 1.50 & 1.49& 0.04& 0.94& 32&0.061& $-9.7$ \\
3.2 &  1.46  &1.44 & 0.09&0.90 & 47 & 0.036& $-10.3$ \\
3.4 & 1.40 & 1.39 &0.12&0.86& 52 &0.035& $-10.5$  \\
\hline\noalign{\smallskip}
\end{tabular}
\end{center}
\end{table}

To explore the initial parameter space for the progenitors of UCXBs
formed by anomalous magnetic braking, we have calculated the evolution
of a large number of IMXBs in the $P_{\rm orb,i}$\,--\,$M_{\rm d,i}$
plane, which was divided into 10 $\times$ 13 discrete grids. The IMXBs
that can evolve into an ultra-short orbital period of less than 60 min
in 13 Gyr are indicated as filled circles in Figure 4. The right
boundary at 3.4 $\rm M_{\odot}$ indicates the maximum mass beyond
which mass transfer from the donor star would be dynamically
unstable. Such dynamical mass transfer would most likely
give rise to
  the spiral-in of the neutron star inside the donor star and the
subsequent merger of the system
  \citep{pods02}. The solid curve shows the bifurcation
period of IMXBs for different donor-star masses. IMXBs with longer
orbital periods evolve into binary millisecond pulsars with a He WD
and a long orbital period. It is impossible to evolve into UCXBs in
the Hubble time for the IMXBs below the bottom boundary. The exact
location in this parameter space depends, however, on the surface
magnetic field and the wind driving efficiency of the donor star.

\begin{figure}
\centering
\includegraphics[width=\linewidth,trim={30 0 80 0},clip]{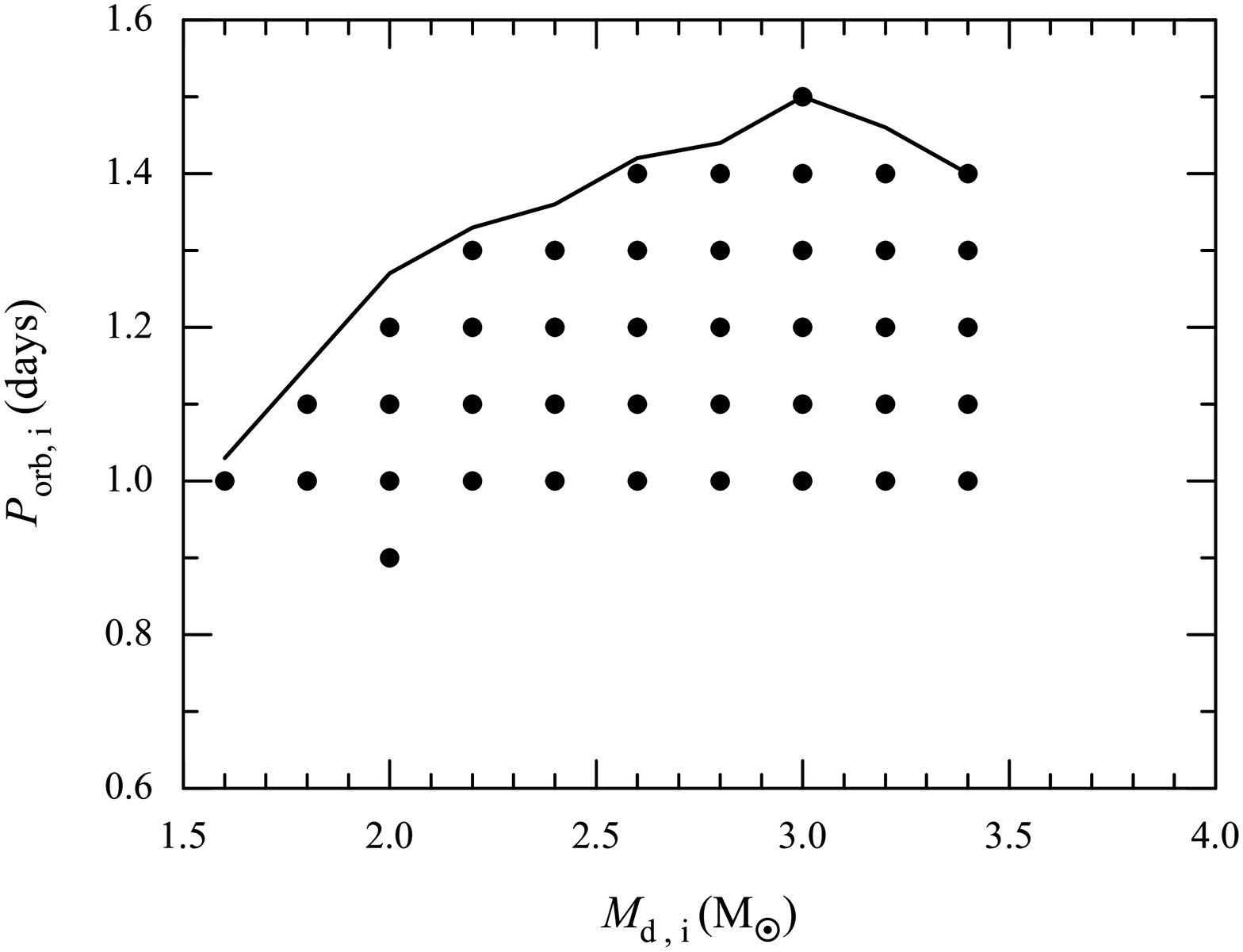}
\caption{Parameter space distribution of the progenitor systems of
  UCXBs formed by the anomalous magnetic braking evolutionary channel
  (within 13 Gyr, with $f=10^{-3}$) in the initial orbital vs.\ initial
  donor-star mass diagram. The solid curve represents the bifurcation
  period of IMXBs for different donor-star masses.} \label{fig:orbmass}
\end{figure}

\begin{figure*}
\centering
\includegraphics[width=2.0\columnwidth]{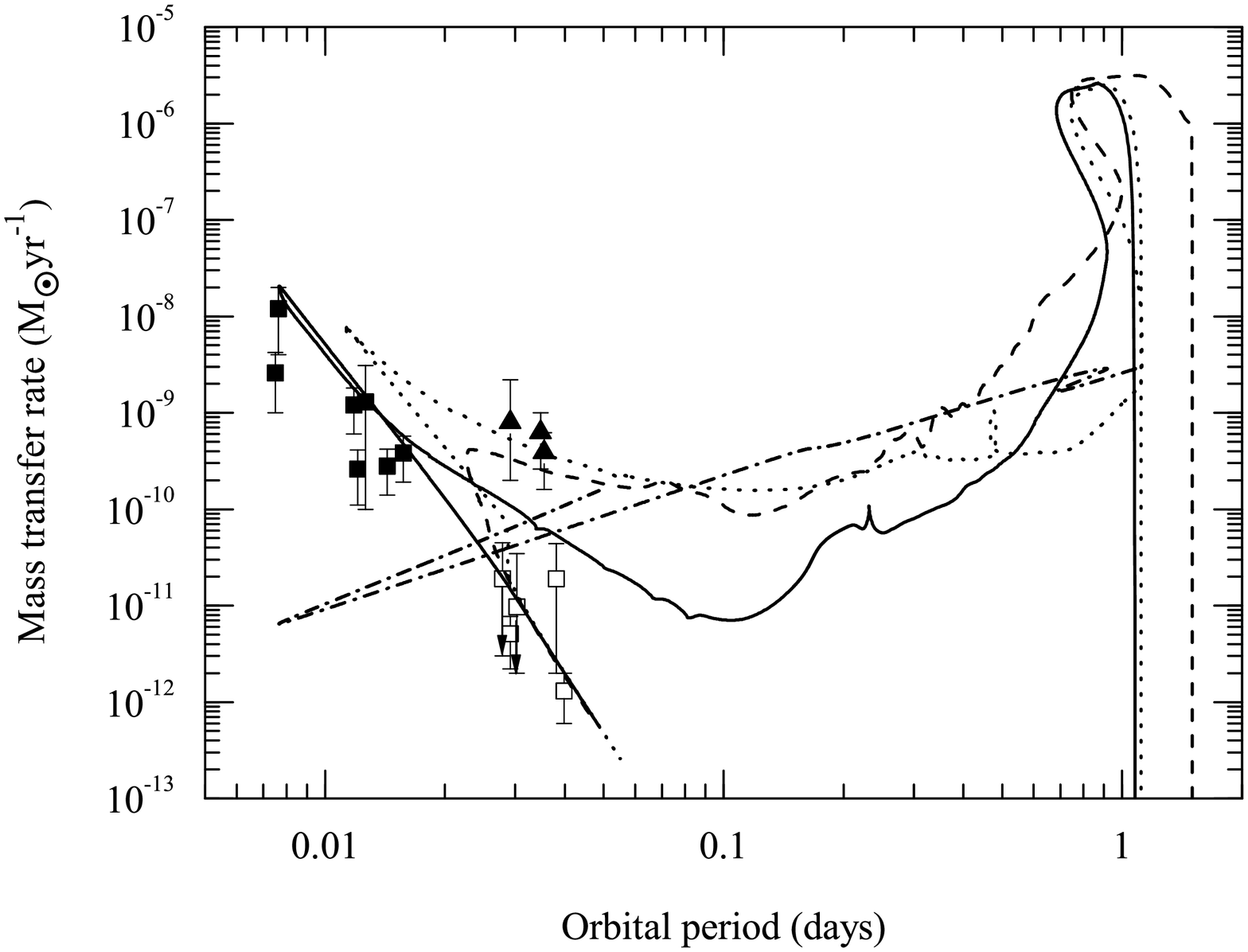}
\caption{Evolution of IMXBs with an initial donor-star mass of $3
  M_{\odot}$ and a period just below the respective bifurcation period
  in the mass-transfer rate vs.\ orbital period diagram. The solid,
  dashed, and dotted curves represent the evolutionary tracks in the
  anomalous magnetic braking model with $f=10^{-5}$ (with a
  bifurcation period of 1.08 d), $f=10^{-3}$ (with a bifurcation
  period of 1.5 d), and the standard magnetic braking with $f=0$ (with
  a bifurcation period of 1.12 d), respectively. The solid squares,
  solid triangles, and open squares denote seven persistent sources
  with a short orbital period, three persistent sources with a
  relatively long orbital period, and five transient sources,
  respectively. The dash-dotted curve corresponds to the critical
  mass-transfer rate for the model track shown as a solid curve, below which
  systems should appear as transients.} \label{fig:orbmass}
\end{figure*}

Table 2 lists the orbital periods and average mass-transfer rates of
ten persistent sources and five transient sources. The mean
mass-transfer rates can be derived from the time-averaged X-ray
luminosity when the mass and the radius of the NS are assumed to be 1.4
$\rm M_{\odot}$, and 11.5 km, respectively \citep{cart13,hein13}. To compare
our models with observations, we plot the evolution of an IMXB with
an initial donor-star mass of $3 M_{\odot}$ and an initial
orbital period of 1.5 days in Figure 5. The
evolutionary results of the standard magnetic braking ($\gamma=4$) given by
\cite{verb81} and \cite{rapp83} are also shown in this figure (here we assume that the standard magnetic braking
model works when the donor star develops a convective envelope). It is clear that our simulated
result is consistent with the observed data for three persistent
sources and four transient sources. Moreover, the mass-transfer rates
of the five transient sources are clearly lower than the critical
mass-transfer rate for the disk-instability model \citep{para96,dubu99}
\begin{eqnarray}
\dot{M}_{\rm cr} \simeq
3.2\times10^{-9}\left(\frac{M_{\rm NS}}{1.4M_{\odot}}\right)^{0.5}
\left(\frac{M_{\rm d}}{1M_{\odot}}\right)^{-0.2} \nonumber\\
\left(\frac{P_{\rm orb}}{1 \rm d}\right)^{1.4}M_{\odot}{\rm yr}^{-1}.
\end{eqnarray}
Below this critical rate, the accretion disk would experience
thermal/viscous instabilities, where the accreting NSs are observed as
transient X-ray sources with short-lived outbursts separated by long
phases of quiescence. According to our simulations, five of the
transient sources should be in the period-increasing phase, which in
principle might be testable by timing observation in the future.

\subsection{Wind-driving efficiency $f = 10^{-5}$}

To form UCXBs with an ultra-short period ($\la$ 20 min), we calculated
the evolution of some IMXBs adopting a relatively small wind-driving
efficiency $f = 10^{-5}$. According to equation (10), the angular
momentum loss rate in this case is approximately the same as the one
with $B_{\rm s}$ = 100 G and $f = 10^{-3}$.  However, the evolution of
the donor stars is quite dramatically different. In Figure 6, we plot
the evolutionary tracks of IMXB with a donor-star mass of $3~\rm
M_{\odot}$ and an initial orbital period of 1.08 days. The donor star
begins to fill its Roche lobe on the main sequence (when $X_{\rm
  c}=0.247$, and $Y_{\rm c}=0.737$).  Due to the low irradiation wind
loss, at $t$ = 9 Gyr, the donor star develops a He core of 0.105 $\rm
M_{\odot}$ (the maximum mass of the He core is $\sim 0.03 \rm
M_{\odot}$ when $f = 10^{-3}$). After the He core overflows its Roche
lobe, it triggers a phase of relatively high mass transfer at a rate
of $10^{-9} - 10^{-8}\, \rm M_{\odot}yr^{-1}$.  When $t$ = 9.45 Gyr,
the binary reaches an ultra-short period of 11 min; at this point the
mass-transfer rate attains a maximum of $2.1\times 10^{-8}\, \rm
M_{\odot}yr^{-1}$. Subsequently, the UCXB begins to widen its
orbit. The timescale in the period-increasing and period-decreasing
stage are $\sim$ 0.1 and $\sim$ 2 Gyr, respectively.

The evolutionary sequences using the standard magnetic braking model
with $f=0$ is also shown in Figure 6.  The calculation shows that the standard magnetic braking formalism has a
much higher efficiency than anomalous magnetic braking in extracting
angular momentum, resulting in a high mass-transfer rate at an earlier
stage. In this case, the UCXB reaches its minimum period of 16 min at
an age of 2.39 Gyr, preceding the anomalous magnetic braking by 7 Gyr.
Therefore, the standard magnetic braking may overestimate the
angular-momentum loss rate. This problem had already been noted in
observations of rapidly rotating stars with a spin period less than
two to five days in young open clusters \citep{quel98,andr03}, in
which the timescale of angular-momentum loss appears to be
approximately two orders of magnitude longer than the value predicted
in the standard magnetic braking model. Some observations of coronal and
chromospheric activity imply that the resulting magnetic braking
attains a maximum at an orbital period less than 3 d
\citep{vilh82,vilh83}. Figure 7 presents the comparison of the
angular-momentum loss rate of the three cases including the strong ($f =
10^{-3}$) and weak ($f = 10^{-5}$) anomalous magnetic braking, and the
standard magnetic braking ($f = 0$). When the donor-star mass
decreases to $0.8~\rm M_{\odot}$, and $0.4~\rm M_{\odot}$, the
standard magnetic braking begins to exceed the weak anomalous magnetic
braking, and the strong one, respectively. For a donor star less
than $0.5~\rm M_{\odot}$, the angular momentum loss rate by the
standard magnetic braking is one order of magnitude greater than that
of the weak anomalous magnetic braking. To interpret the period gap of
cataclysmic variables, \cite{verb84} proposed that the donor stars
need to have a magnetic field of $\ga$ 100 G and a wind of
$10^{-10}~\rm M_{\odot}\,yr^{-1}$, which is compatible with
our result.

Figure 8 shows that the initial parameter space for UCXBs formed by
weak magnetic braking is much smaller than that for strong magnetic
braking (see also Figure 4). Except for initial donor stars in the
range of $1.6-2.0~\rm M_{\odot}$, the most massive IMXBs all have the
same bifurcation period of 1.08 d. Compared to the weak anomalous
magnetic braking, standard magnetic braking has a relatively long
bifurcation period (1.10 - 1.12 d for donor stars of $2.4-3.4~\rm
M_{\odot}$) and a slightly smaller initial parameter space.

Figure 5 also presents the predicted relation between the orbital
period and the mass-transfer rate for the anomalous magnetic braking
with $f=10^{-5}$, and the standard magnetic braking with $f=0$. In the
final stage, the magnetic braking ceases in all cases and
gravitational radiation becomes the dominant driving mechanism; so the
evolutionary tracks are similar in the three cases. It is clear that the
anomalous magnetic braking scenario with a small wind-driving
efficiency can fit four persistent sources and four transient
sources. However, this case cannot produce the long-period UCXBs with
a relatively high luminosity. The standard magnetic braking model can
fit the observed data of three persistent sources with long periods and
four transient sources, which is similar to the anomalous magnetic
braking with $f=10^{-3}$.

\begin{figure*}
\centering
\begin{tabular}{ccc}
\includegraphics[width=0.33\textwidth,trim={0 0 0 0},clip]{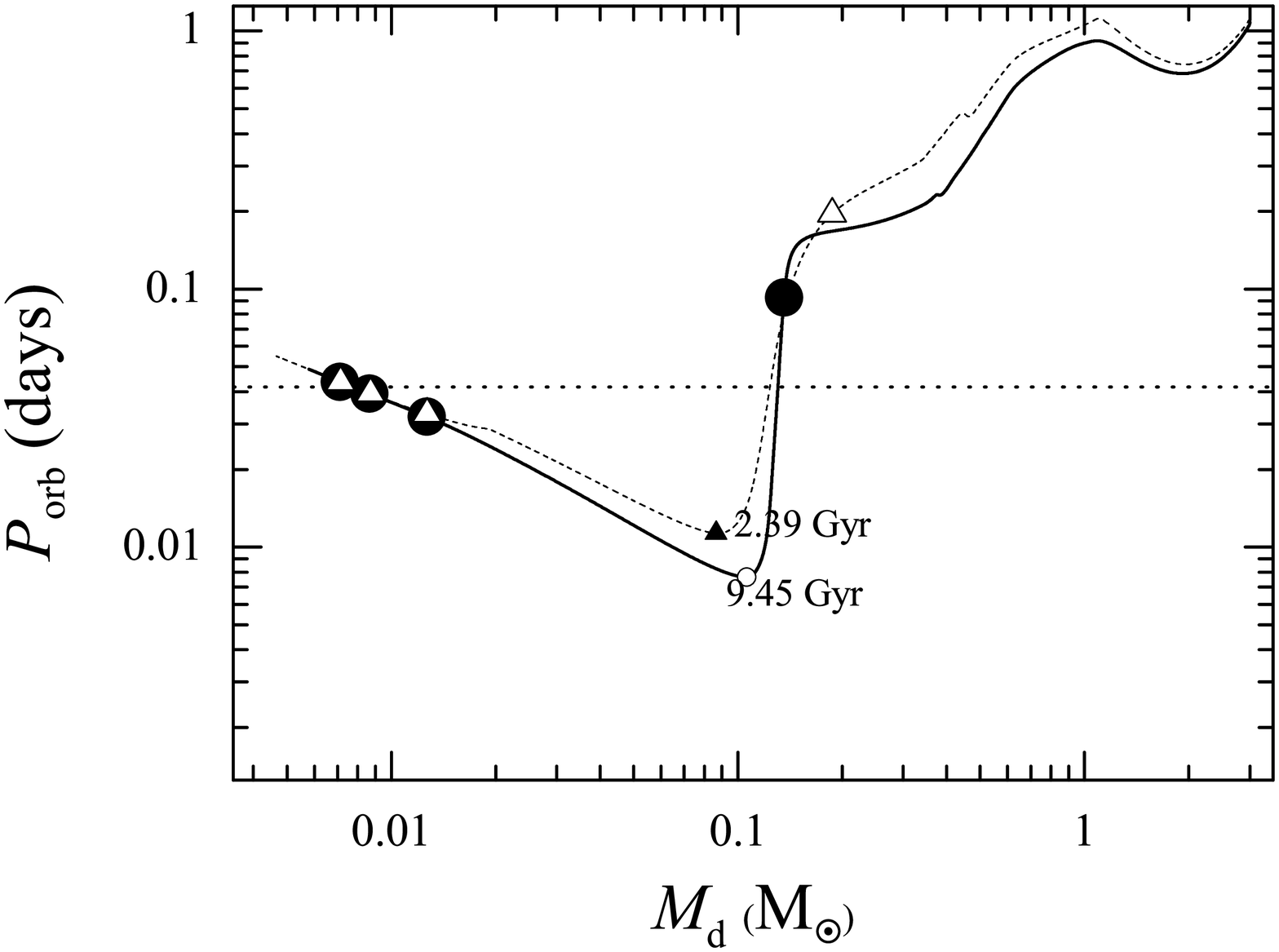} &    \includegraphics[width=0.33\textwidth,trim={0 0 0 0},clip]{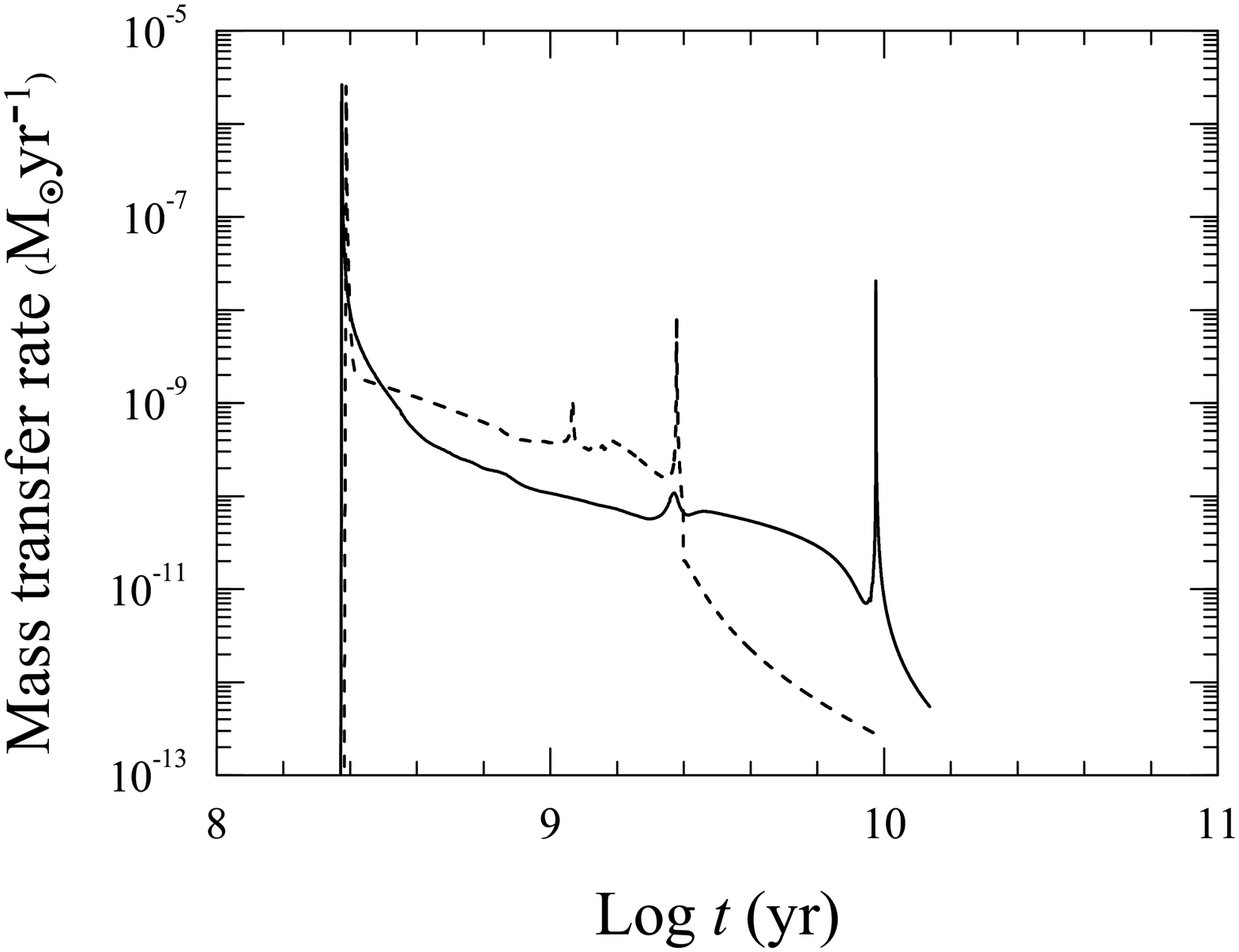} &\includegraphics[width=0.33\textwidth,trim={0 0 0 0},clip]{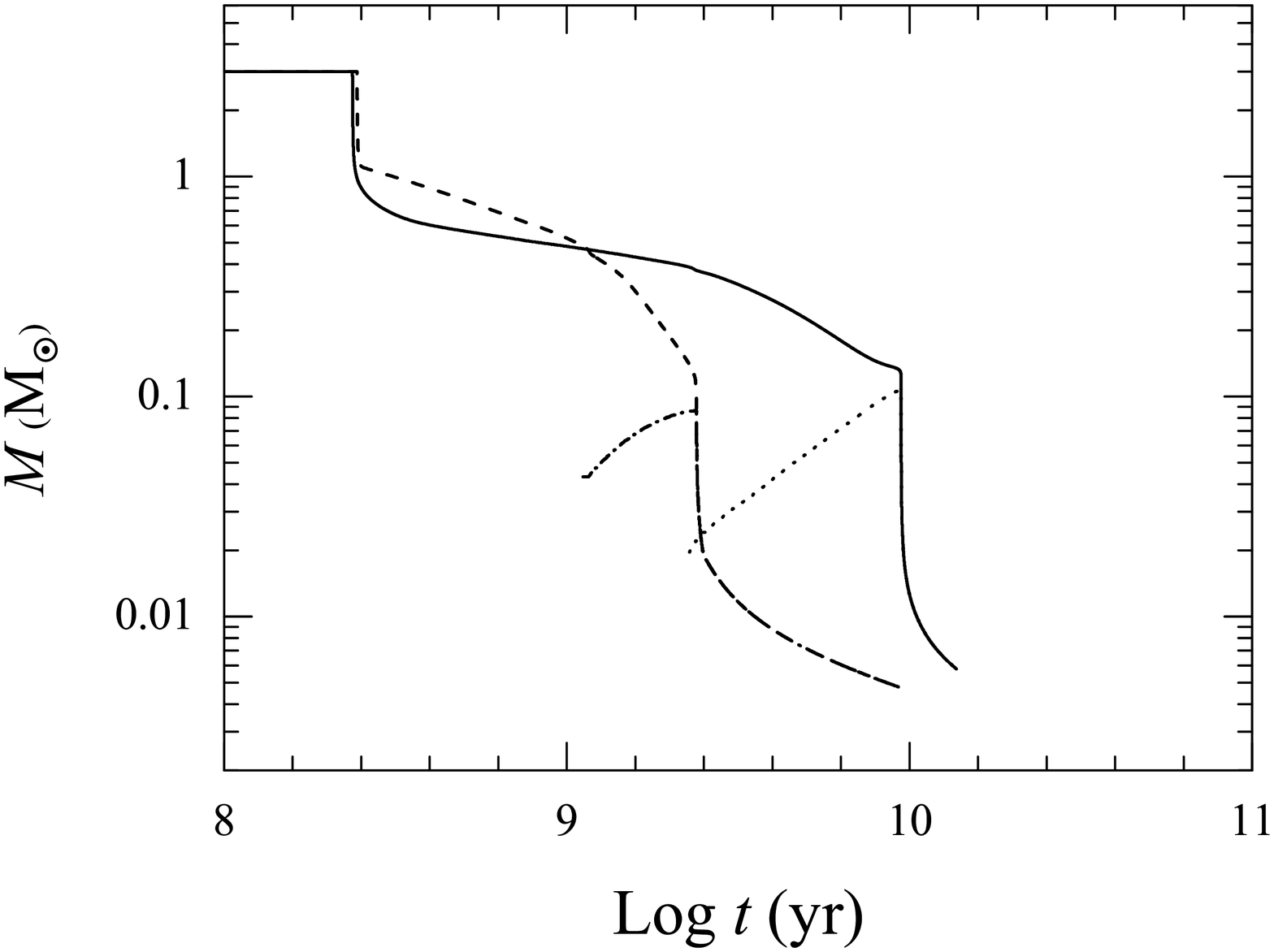} \\
\end{tabular}
\caption{\label{fig:pulses}Evolutionary tracks of an IMXB with a donor
  star mass of $3~\rm M_{\odot}$ for both the anomalous magnetic
  braking model with $f=10^{-5}$ (with a bifurcation period of 1.08
  days) and the standard magnetic braking model with $f=0$ (with a
  bifurcation period of 1.12 days). The solid and dashed curves represent
  the evolutionary tracks of the anomalous magnetic braking and the
  standard magnetic braking, respectively. The horizontal dotted line in the left panel denotes
  the maximum period (1 hour) defining UCXBs. Left panel: orbital period as function
of donor mass; the solid circles along
  the solid curve
  indicate ages of 9, 10, 11, 12 Gyr, respectively; the open triangles along the
  dashed curve
  represent ages of 2, 3, 4, 5 Gyr, respectively. Middle panel:
  mass-transfer rate. Right panel: the mass of the donor star and the He core (the
  dotted and dash-dotted curves represent the anomalous magnetic
  braking and the standard magnetic braking, respectively).}
\end{figure*}

\begin{figure}
\centering
\includegraphics[width=\linewidth,trim={30 0 80 0},clip]{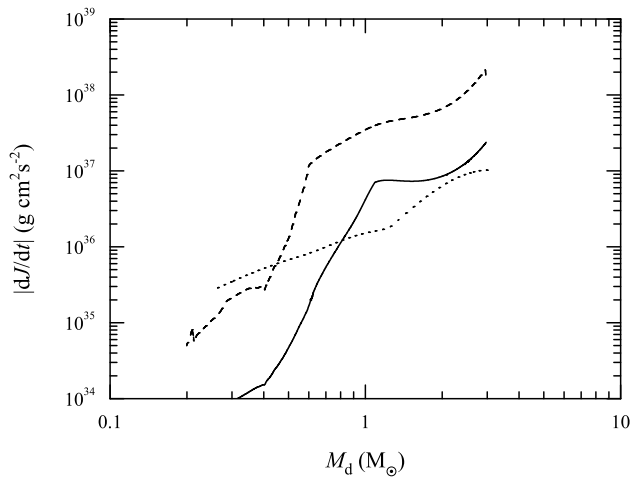}
\caption{Angular-momentum loss rate as the function of the donor-star
  mass for an IMXB with a $3 {\rm M}_{\odot}$ donor star and an initial
  orbital period of 1.0 d. The solid, dashed, and dotted curves
  correspond to the anomalous magnetic braking with $f=10^{-5}$,
  $f=10^{-3}$, and the standard magnetic braking with $f=0$,
  respectively. } \label{fig:orbmass}
\end{figure}

\begin{figure}
\centering
\includegraphics[width=\linewidth,trim={30 0 80 0},clip]{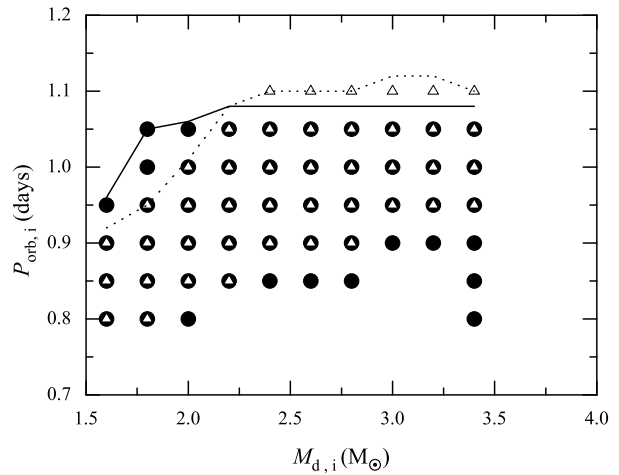}
\caption{Similar to Figure 4, the solid circles and open triangles
  denote IMXBs evolved by the anomalous magnetic braking with
  $f=10^{-5}$ and the standard magnetic braking with $f=0$,
  respectively. The solid and dotted curves represent the bifurcation
  period of IMXBs evolved by the anomalous magnetic braking and the
  standard magnetic braking, respectively. } \label{fig:orbmass}
\end{figure}

\section{Discussion}

It is difficult to form UCXBs with an orbital period less than 10 min
in our scenario. In this section, we first analyze the factors
influencing the minimum period.  In principle, the orbital evolution
of X-ray binaries depends on the angular-momentum loss and the
mass-transfer process. The rate of change of the orbital period satisfies
the relation
\begin{equation}
\frac{\dot{P}_{\rm orb}}{P_{\rm orb}}=3\frac{\dot{J}}{J}-3\frac{\dot{M}_{\rm d}}{M_{\rm d}}(1-q\alpha)+\frac{\dot{M}_{\rm NS}+\dot{M}_{\rm d}}{M_{\rm NS}+M_{\rm d}},
\end{equation}
where $\alpha=-\dot{M}_{\rm NS}/\dot{M}_{\rm d}$ is the ratio between
the accretion rate of the NS and the mass loss rate of the donor
star. For X-ray binaries with an orbital period less than 2 hours,
angular-momentum loss via gravitational radiation surpasses that due
to magnetic braking \citep{nels03}. Then, ignoring the angular
momentum carried away by mass loss, $\dot{J}=\dot{J}_{\rm GR}$. In
equation (12), $\dot{P}_{\rm orb} = 0$ yields the minimum
period. Therefore, the minimum period can be written as
\begin{equation}
P_{\rm orb,min}=0.066\frac{m_{\rm NS}^{3/8}m_{\rm d}^{3/4}(m_{\rm NS}+m_{\rm d})^{-1/8}}{(1-q\alpha)^{3/8}\mid\dot{m}_{\rm d}\mid^{3/8}}\, \rm  min,
\end{equation}
where $m_{\rm NS}$ and $m_{\rm d}$ are $M_{\rm NS}$, and $M_{\rm d}$
in units of solar masses, and $\dot{m}_{\rm d}$ is $\dot{M}_{\rm d}$ in
units of $\rm M_{\odot}yr^{-1}$.

In Figure 1, the high magnetic field leads to efficient angular
momentum loss and rapid mass transfer. For low magnetic fields, with
mass transfer occurring on a rather long timescale, the system
develops a relatively tight orbit for the same donor-star
mass. Because of the angular momentum loss induced by the strong
  gravitational radiation, a relatively high mass-transfer rate is
  expected. When $f = 10^{-5}$, the massive He core that develops from
  the donor star leads to a more compact orbit, again resulting in a much
  higher mass-transfer rate.

In Figure 9, we show the relation between the minimum period and the
mass-transfer rate for three different donor-star masses and a
constant NS mass of $1.5~\rm M_{\odot}$. If the observed sources reach
the minimum period during mass transfer, they should be
  located above the relevant curves at the present time (unless they
experience a detached stage before the minimum period). According to
this figure, the donor-star masses can be constrained to be
near $\sim$ 0.1, $\sim$ 0.05, and $\sim$ 0.01 $\rm M_{\odot}$
for the three long-period persistent sources, seven short period
persistent sources, and five transient sources, respectively. In
Figure 5, our simulated results show that the relevant donor-star
masses are in the range of $0.099-0.116$, $0.0358-0.108$, and
$0.010-0.016$ $\rm M_{\odot}$, respectively. These inferred values are
in approximate agreement with the calculated results.

\begin{table}
\begin{center}
\caption{Average mass-transfer rates and orbital periods of some UCXBs\label{tbl-2}}
\begin{tabular}{cccc}
\hline\hline\noalign{\smallskip}
Source & $P_{\rm orb}$ &  $<\dot{M_{\rm tr}}>$ &References\\
       & (minutes) &  $\rm M_{\odot}yr^{-1}$              &            \\
\hline\noalign{\smallskip}
Persistent sources \\
\hline\noalign{\smallskip}
4U 1728-34 & 10.8 &  $2.6\pm1.6\times 10^{-9}$ & 1 \\
4U 1820-303 & 11 &  $1.2\pm0.8\times 10^{-8}$ & 2,3 \\
4U 0513-40 & 17 &  $1.2\pm0.6\times 10^{-9}$ & 2,4 \\
2S 0918-549 & 17.4 &  $2.6\pm1.5\times 10^{-10}$ & 5,6 \\
4U 1543-624 & 18.2 &  $1.3^{+1.8}_{-1.2}\times 10^{-9}$ & 7 \\
4U 1850-087 & 20.6 &  $2.8\pm1.4\times 10^{-10}$ & 2,8 \\
M15 X-2 & 22.6 &  $3.8\pm1.9\times 10^{-10}$ & 2,9 \\
4U 1627-67 & 42 &  $8^{+14}_{-6}\times 10^{-10}$ & 10 \\
4U 1916-053 & 50 &  $6.3\pm3.7\times 10^{-10}$ & 11,12,13 \\
4U 0614+091 & 51 & $3.9\pm2.3\times 10^{-10}$ & 14,15 \\
\hline\noalign{\smallskip}
Transient sources \\
\hline\noalign{\smallskip}
XTE J1807-294 & 40.1 &  $<1.9^{+2.6}_{-1.6}\times 10^{-11}$ & 16,17 \\
XTE J1751-305 & 42 &  $5.1^{+2.6}_{-2.9}\times 10^{-12}$ & 18,19 \\
XTE J0929-314 & 43.6 &  $<9.7^{+25}_{-7.7}\times 10^{-11}$ & 16,20 \\
Swift J1756.9-2508 & 54.7 &  $1.9^{+2.5}_{-1.7}\times 10^{-11}$ & 21 \\
NGC 6440 X-2 & 57.3 &  $1.3\pm 0.7\times 10^{-12}$ & 2,22 \\
\hline\noalign{\smallskip}
\end{tabular}
\tablenotetext{}{References. (1) \cite{gall10}; (2) \cite{harr10}; (3) \cite{stel87}; (4) \cite{zure09}; (5) \cite{int05}; (6) \cite{zhon11}; (7) \cite{wang04}; (8) \cite{home96}; (9) \cite{dieb05};
(10) \cite{chak98}; (11) \cite{yosh93}; (12) \cite{walt82}; (13) \cite{iari15}; (14) \cite{bran92}; (15) \cite{shah08};
(16) \cite{gall06}; (17) \cite{mark03}; (18) \cite{papi08};
(19) \cite{mark02}; (20) \cite{gall02}; (21) \cite{krim07};  (22) \cite{alta10}.}
\end{center}
\end{table}

\begin{figure}
\centering
\includegraphics[width=\linewidth,trim={30 0 80 0},clip]{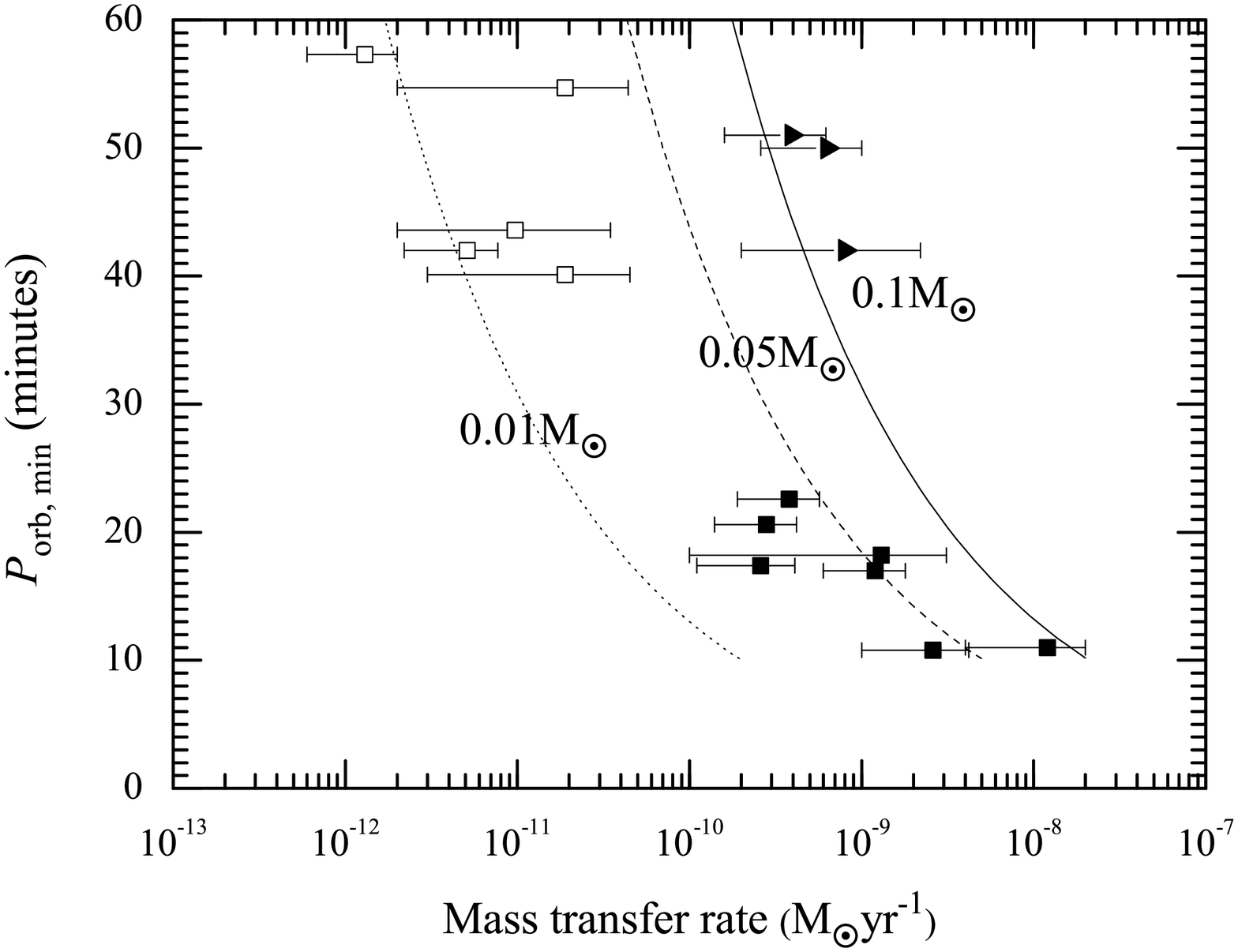}
\caption{The minimum orbital period of UCXBs as a function of the
  mass-transfer rate if the orbital decay is only driven by
  gravitational radiation. The solid, dashed, and dotted curves correspond
  to donor star-masses of 0.1, 0.05, and 0.01 $\rm M_{\odot}$,
  respectively. The other symbols are similar to these in Figure
  5.} \label{fig:orbmass}
\end{figure}

Statistically, only a small fraction of A/B stars (about $ 5\%$) stars
have anomalously strong magnetic fields and appear as Ap/Bp stars
\citep{land82,shor02}. However, the populations formed via our
evolutionary route should appear as UCXBs for a much longer time than
in the circumbinary disk model \citep{ma09b}. The former spend 0.2 and
2.0 Gyr in the period-decreasing and period-increasing UXCB phases,
while the lifetime of the latter is only 0.01 Gyr (there is no
period-increasing stage in that model). Therefore, for the same birth rate of
UCXBs, the number of observable systems formed by our scenario is approximately one order of magnitude
larger than that in the circumbinary disk model. Compared to our
evolutionary channel, the advantage of the circumbinary disk model is
that it can drive LMXBs to an ultra-short period as short as 6 min; however,
such a short-period UCXB has not yet been detected.

The irradiation wind-driving efficiency or magnetic field play an
important role in influencing the formation of UCXBs such as the
minimum period and the mass-transfer rate. A small $f = 10^{-5}$ or
weak magnetic field ($B_{\rm}$ = 100 G) can lead to
minimum periods of 11 min, while a large $f = 10^{-3}$ may be applicable to
the long-period persistent sources. Because the magnetic fields of the
Ap/Bp stars are in the range of 100 - 10000 G, the anomalous magnetic
braking scenario provides a reasonable evolutionary channel towards
various UCXBs.

\section{Conclusions}
It is generally thought that UCXBs with orbital periods $< 50$ min
cannot form by a modified magnetic braking mechanism \citep{sluy05b}
unless a circumbinary disk around L/IMXBs exists \citep{ma09b}. In
this work, we propose an alternative evolutionary route towards
forming UCXBs. Some IMXBs, which contain Ap/Bp stars with an anomalously
strong magnetic field ($100 - 10000$ G) may be able to produce UCXBs
by anomalous magnetic braking with irradiation processes taken into
account.  To test the possibility of this new evolutionary channel for
UCXBs, we have performed evolutionary calculations for a large
number of IMXBs. Our main results are summarized as follows.

1. For donor stars with a 1000 G magnetic field, anomalous magnetic
braking can evolve IMXBs into UCXBs with a minimum orbital period of 11
or 32 min (in our example calculations) when the irradiation
wind-driving efficiency is $f = 10^{-5}$ and $10^{-3}$,
respectively. A smaller wind-driving efficiency or weaker magnetic
field would give rise to a shorter orbital period.

2. Three long-period persistent UCXBs with a relatively high
mass-transfer rate of $\sim 10^{-9} \,\rm M_{\odot}yr^{-1}$ can be
produced by anomalous magnetic braking with a high irradiation
wind-driving efficiency ($10^{-3}$) or a strong magnetic field. However, seven
short-period persistent UCXBs favour a low wind-driving
efficiency or weak magnetic field.

3. The evolutionary timescale of our simulated UCXBs in the
period-increasing phase is generally much longer than that in the
orbit-decaying phase. In the example presented in the paper, the
evolutionary timescales in these two phases are $\sim$ 2 and 0.1 Gyr,
respectively. Therefore, our scenario can produce a relatively high
birth rates despite of a low fraction (5\%) of Ap/Bp stars among
intermediate-mass stars.

4.  For intermediate-mass donor stars in the range of $1.6 - 3.4$ $\rm
M_{\odot}$, IMXBs with orbital periods in the range of $0.9 - 1.5$ d (for
$f = 10^{-3}$; when $f = 10^{-5}$, the orbital period range changes to $0.8 -
1.08$ d), the corresponding IMXBs can evolve into UCXBs.

5. Because H is almost exhausted in the centre, IMXBs near the bifurcation
period can reach the shortest orbital periods for a specific donor-star
mass, similar to the conclusions drawn by \cite{pods02} and
\cite{sluy05a}.

6. Our calculations indicate that the standard magnetic
  braking model may overestimate the angular-momentum loss
  rate for low-mass ($<0.8~\rm M_{\odot}$) donor stars. Though the
  angular momentum loss rate by weak anomalous magnetic braking
  ($f=10^{-5}$) is approximately $1-2$ orders of magnitude lower than
  that in the standard magnetic braking model, it can still produce a much
  smaller minimum orbital period of 11 min, and has a much wider
  initial parameter space for forming UCXBs.

\acknowledgments {We are grateful to the anonymous referee for very
  helpful and useful suggestions.  This work was partly supported by
  the National Science Foundation of China (under grant number
  11573016), Program for Innovative Research Team (in Science and
  Technology) in University of Henan Province Province, and China
  Scholarship Council.}


\end{document}